\documentclass{article}

\usepackage{arxiv}

\usepackage[utf8]{inputenc} % allow utf-8 input
\usepackage[T1]{fontenc}    % use 8-bit T1 fonts
\usepackage{url}            % simple URL typesetting
\usepackage{booktabs}       % professional-quality tables
\usepackage{amsfonts}       % blackboard math symbols
\usepackage{nicefrac}       % compact symbols for 1/2, etc.
\usepackage{microtype}      % microtypography
\usepackage{lipsum}		% Can be removed after putting your text content
\usepackage{authblk}    % Can help for add authors affiliations 
\usepackage{graphicx}
\usepackage{doi}

\usepackage{cite}
\usepackage{amsmath,amssymb,amsfonts}
\usepackage{algorithmicx}
\usepackage{algpseudocode}
\usepackage[linesnumbered,ruled,vlined]{algorithm2e}
\usepackage{textcomp}

\usepackage{ifpdf}
\usepackage{pslatex}
\usepackage{xcolor}
\usepackage{textcomp}
\usepackage{placeins}
\usepackage{eqparbox}
\usepackage{stfloats}
\usepackage{multicol}
\usepackage{subfigure}
\usepackage{caption,setspace}
\usepackage{float}
\usepackage{amsthm}
\usepackage{comment}
\usepackage{color, soul}
\usepackage{enumerate}

\usepackage{array}
\usepackage{balance}
\usepackage{color}
\usepackage{makecell}

\title{\emph{Remote Detection of Applications for Improved Beam Tracking in mmWave/sub-THz 5G/6G Systems}}

\author[1,2]{Alexander Shurakov}
\author[1,2]{Margarita Ershova}
\author[3]{Abdukodir Khakimov}
\author[1,2,4]{Anatoliy Prikhodko}
\author[3]{Evgeny Mokrov}
\author[3]{Vyacheslav Begishev}
\author[1,2]{Galina Chulkova}
\author[1,$\star$]{Yevgeni Koucheryavy}
\author[1,2,4]{Gregory Gol{'}tsman}

\affil[1]{HSE University, Moscow, Russia}
\affil[2]{Moscow Pedagogical State University, Moscow, Russia}
\affil[3]{Peoples' Friendship University of Russia (RUDN University), Moscow, Russia}
\affil[4]{Russian Quantum Center, Skolkovo, Russia}
\affil[$\star$]{email: ykoucheryavy@hse.ru}

\renewcommand{\shorttitle}{Shurakov et al.}

\begin{document}
\maketitle

\begin{abstract}
Beam tracking is an essential functionality of millimeter wave (mmWave, 30--100 GHz) and sub-terahertz (sub-THz, 100--300 GHz) 5G/6G systems. It operates by performing antenna sweeping at both base station (BS) and user equipment (UE) sides using the Synchronization Signal Blocks (SSB). The optimal frequency of beam tracking events is not specified by 3GPP standards and heavily depends on the micromobility properties of the applications currently utilized by the user. In absence of explicit signalling for the type of application at the air interface, in this paper, we propose a way to remotely detect it at the BS side based on the received signal strength pattern. To this aim, we first perform a multi-stage measurement campaign at 156~GHz, belonging to the sub-THz band, to obtain the received signal strength traces of popular smartphone applications. Then, we proceed applying conventional statistical Mann-Whitney tests and various machine learning (ML) based classification techniques to discriminate applications remotely. Our results show that Mann-Whitney test can be used to differentiate between fast and slow application classes with a confidence of 0.95 inducing class detection delay on the order of 1 s after application initialization. With the same time budget, random forest classifiers can differentiate between applications with fast and slow micromobility with 80\% accuracy using received signal strength metric only. The accuracy of detecting a specific application however is lower, reaching 60\%. By utilizing the proposed technique one can estimate the optimal values of the beam tracking intervals without adding additional signalling to the air interface.
\end{abstract}

% keywords can be removed

\section{Introduction}
The need for more bandwidth at the air interface has led to the adoption of millimeter wave (mmWave, 30--100 GHz) band in 5G New Radio (NR) systems \cite{storck2020survey}. Now, the aim is to utilize even higher sub-terahertz (sub-THz, 100--300 GHz) band for prospective 6G cellular systems \cite{polese2020toward,wang2021key}. The small wavelength in these bands limits the radiation power of a single antenna element and naturally requires the use of massive antenna arrays operating in the beamforming mode to extend the coverage of base stations (BS) \cite{pant2023thz,nissanov2023multi}. Moreover, the thrifty distribution of radio signals with the aid of narrow-beam antenna systems is conditioned by the fact that the existing solutions in mm-Wave/sub-THz integrated transceivers and state-of-the-art semiconductor components ensure less than a watt of signal power available at the air interface \cite{du2023sige}.

% Beamforming: frequency is not set
Beam tracking is an essential feature of mmWave/sub-THz 5G/6G systems distinguishing them from previous generations of cellular systems \cite{giordani2018tutorial}. The goal of beam tracking is to ensure that the BS and user equipment (UE) beams are well aligned at all the time instants even when UE is mobile. This functionality is currently implemented in 5G NR by regularly performing antenna sweeps using Synchronization Signal Blocks (SSB) at both BS and UE sides for all the active users in the cell \cite{nrmcs}. However, the standard does not specify precisely the frequency of beam tracking. When this frequency is rather high while UE is stationary, a large set of time and frequency resources are wasted that can otherwise be utilized for data transmission. Thus, optimizing the beam tracing frequency is critical for enhancing the performance of systems operating in the beamforming mode.

% Micromobility

One of the most important factors affecting beam alignment in mmWave/sub-THz systems is micromobility \cite{petrov2018effect}. Micromobility refers to small displacements and rotations of the UE in hands of a user even when he/she is in stationary state. As opposed to blockage \cite{shurakov2023dynamic}, beam misalignment due to micromobility happens at the smaller timescales that is compared to the range of time beam tracking time intervals \cite{nrmcs}. As shown in \cite{petrov2020capacity}, micromobility may lead to temporal outages when the SSB frequency is not sufficient. This phenomenon, initially predicted theoretically, has been recently quantified in \cite{ichkov2021millimeter,stepanov2021accuracy} for 5G NR 
 systems operating in the mmWave band, where notable performance degradation has been reported.

% Dependence on application

It has been recently shown that micromobility pattern is mainly dictated by the currently utilized application \cite{stepanov2021statistical}. The analysis of micromobility patterns demonstrated that different applications are characterized by principally different properties including the speed, time to misalignment, etc. In absence of explicit signalling about the currently utilized application type at the UEs, BS cannot make a guided decision on the optimal beam tracing interval to minimize the usage of resources while still ensuring outage-free operation. However, as micromobility patterns of different applications statistically vary \cite{stepanov2021accuracy}, BS may implicitly deduce the application type remotely by using classification techniques based on the received signal strength.

% What we do

In this paper, we will develop a simple approach for remote detection of applications currently utilized at UEs based on their micromobility patterns. To this aim, we will first perform a multi-stage measurement campaign in the sub-THz frequency band at 156 GHz. We measure and analyze the statistical properties of the received signal strength of four typical applications including voice calling, video watching, VR watching, and gaming. Having derived the stochastic received signal strength series, we will then proceed proposing and comparing classic statistical tests and machine learning (ML) techniques for reliable classification of applications into two categories characterized by fast and slow micromobility behavior which is sufficient for estimating the optimal beam tracking interval characterized by outage-free operation. We also apply the developed tests for identifying a particular class of applications and show that the received signal strength only is insufficient for this purpose. %We report the typical inter-synchronization times resulting in the best trade-off between outage and resource utilization.

% Contributions

The main contributions of our study are:
\begin{itemize}
    \item{real measurements and analysis of the received signal strength patterns at 156 GHz under micromobility impairments of typical smartphone applications including video, voice, VR, and gaming;}
    \item{a method and algorithm for remote detection of the application class and type based on Mann-Whitney test and ML techniques including trees and random forests for the received signal strength just after the beam tracking time instant;}
%    \item{a method for remote detection of the application class and type based on Mann-Whitney test for the slope difference in the received signal strength just after the beam tracking time instant;}
    \item{results showing that Mann-Whitney test allows to differentiate between application classes with a $0.95$ confidence limit in just 1 s; random forests outperform other ML-based classifiers (conventional as well as ML-based) providing higher than 80\% of accuracy in more than 5 s.}
\end{itemize}

% Structure

The rest of the paper is organized as follows. We begin in Section \ref{sect:rel} providing an outline of the related work. In Section \ref{sect:meas}, we describe the measurement campaign. Then, in Section \ref{sect:stat}, we report statistical characteristics of the micromobility patterns. Mann-Whitney and ML-based classification techniques are introduced later in Section \ref{sect:stat}. Numerical results are reported in Section \ref{sect:numerical}. The conclusions are provided in the last section.

\section{Related Work}\label{sect:rel}
Micromobility, characterized by rapid and frequent changes in a user location, has a significant impact on systems employing directional antennas. While 5G NR systems utilize directional antenna patterns, their half-power beamwidth (HPBW) is relatively large, limiting the effects of micromobility. However, experimental studies, such as those conducted in \cite{ichkov2021millimeter}, have shown that micromobility can lead to a degradation of received signal strength in 5G NR systems.

As we transit to the sub-THz band and beyond, the dimensions of antenna arrays increase, resulting in narrower HPBWs. This narrower beamwidth can enhance directivity but also exacerbates the impact of user micromobility on link performance. The work in \cite{petrov2018effect} characterized the influence of user micromobility on link performance. The authors employed inertial sensors embedded in a smartphone, including a gyroscope and accelerometer, to model the stochastic trajectories of the imaginary boresight of the beam. Their findings revealed a fundamental trade-off between outage time and spectral efficiency for varying HPBW sizes. In \cite{petrov2020capacity}, the same authors utilized simplified micromobility models based on diffusion processes to evaluate the performance of on-demand and regular beam-tracking procedures in the THz band. Their results demonstrated that regular beam-tracking, currently employed in 5G NR systems, may not be the optimal approach when the number of antenna elements reaches levels typical for 6G THz systems.

In \cite{stepanov2021accuracy}, the authors conducted a micromobility emulation using a laser pointer rigidly attached to a smartphone to represent the beam's boresight. They analyzed four application types: video watching, VR viewing, racing games, and phone calling, finding that each exhibited distinct micromobility characteristics. Their study revealed that applications like video watching and phone calling could tolerate longer intervals between beam-tracking time instants. Based on statistical data from \cite{stepanov2021statistical}, three micromobility models were formulated and parameterized in \cite{stepanov2021accuracy}: (i) a two-dimensional Markov model, (ii) a decomposed one-dimensional Markov model along the Oy and Ox axes, and (iii) a decomposed Brownian motion model along the Ox and Oy axes. These models were compared using the time to outage after beam-tracking, with the two-dimensional model demonstrating the closest approximation to statistical data but also the highest computational complexity due to its thousands of states. Decomposed models, while less accurate, provide a suitable first-order approximation for applications where user actions are not directly controlled by the application, such as video watching, phone calling, and VR viewing (excluding racing games).

Following the characterization of micromobility effects, research has begun to assess its impact on user and system performance and to propose solutions for remote application detection. In \cite{moltchanov2021ergodic}, the authors investigated the influence of micromobility on the outage performance of 6G THz systems equipped with micromobility functionality. Their findings revealed that a tenfold reduction in beam searching time is necessary to achieve a meaningful balance between outage and spectral efficiency. This can be accomplished by employing faster antenna configuration switching. In \cite{sopin2022user}, a joint deployment of mmWave and THz 6G communication systems was explored. By considering both channel- and resource-dependent factors, the authors demonstrated that systems with intermittent connectivity caused by micromobility must rely on highly reliable backup options, such as 4G LTE.

Regarding the classification of applications for optimizing the interval between beam-tracking time instants, \cite{dugaeva2022using} proposed utilizing inertial sensors, such as gyroscopes and accelerometers, to narrow the search space for antenna configurations on both the BS and UE sides. However, the authors demonstrated that the inherent inaccuracies of modern embedded sensors limit the potential gains in beam search time. An alternative approach, presented in \cite{dugaeva2023utilization}, involves detecting the boresight location of the antenna's main lobe. This method can achieve highly accurate application detection, leading to optimal beam-tracking intervals. However, the boresight location metric is not readily available at the BS and UE.

Despite extensive research, there remains a lack of empirical measurements and methodologies to precisely quantify the impact of micromobility on received signal strength immediately following a beam-tracking event.

\begin{table*}[t!]
\vspace{-0mm}
\caption{Summary of micromobility characteristics and modeling insights, adopted from \cite{stepanov2021statistical}}
\label{tab:summary}
\begin{center}\small
\begin{tabular}{|l|l|l|l|l|}
\hline
Characteristics & Video viewing & Phone calling & VR viewing & Racing game \\
\hline\hline
\makecell[l]{Radial \\ symmetry} & Yes & No & No & No \\
\hline
\makecell[l]{Velocity \\ as a function of distance} & Increases, $3-10$%$10-35$
~m/s & Constant, $7$%Increase, $20-35$
~m/s & Increases, $9-13$%$35-50$
~m/s & Decreases, $9-5$%Constant, $30$
~m/s \\
\hline
\makecell[l]{Drift to the origin \\ as a function of distance} & Decreases, $0.17- 0.11$ %Constant, $0.17$ 
& Increases, $0.17-0.3$ %Decreases, $0.17-0.03$%0.08$
& Constant, $0.17$ & Constant, $0.17$ \\
\hline
\makecell[l]{Axes \\ dependence} & Negligible & Moderate & Negligible & Strong \\
\hline
\makecell[l]{Axes increment \\ correlation coefficient} & 0.0 & -0.2 & 0.0 & -0.4 \\
\hline
\makecell[l]{X-axis velocity \\ as a function of distance} & Increases, $1-6$~%$5-20$
m/s & Increases, $3-6$~%$15-25$
m/s & Increases, $6-9$~%Constant, $25$
m/s & Decreases, $7-4$~%Increases, $23-35$
m/s \\
\hline
\makecell[l]{Y-axis velocity \\ as a function of distance} & Increases, $2-8$~%$7-25$
m/s & Increases, $3-5$~%$10-23$
m/s & Increases, $5-8$~%Constant, $10$
m/s & Decreases, $3-2$~%Increases, $20-30$
m/s \\
\hline
\makecell[l]{X-axis drift \\ as a function of distance} & Decreases, $0.17-0.05$%Increases, $0.17-0.22$%Decreases, $0.17-0.1$ 
&Decreases, $0.17-0.13$% Increases, $0.17-0.2$%Decreases, $0.17-0.12$
& Constant, $0.17$ & Increases, $0.13-0.21$%Decreases, $0.2-0.15$%Constant, $0.17$
\\
\hline
\makecell[l]{Y-axis drift \\ as a function of distance} & Increases, $0.17-0.21$%Decreases, $0.17-0.1$%Constant, $0.17$ 
& Increases, $0.17-0.25$%Decreases, $0.17-0.08$%Increases, $0.17-0.24$
& Constant, $0.17$ 
& Decreases, $0.19-0.14$%Increases, $0.15-0.19$%Constant, $0.17$
\\
\hline
\makecell[l]{Markov \\ modeling} & Yes & Yes & Limited & No \\
\hline
\end{tabular}
\end{center}
\vspace{-0mm}
\end{table*}

\section{Micromobility Measurements Campaign}\label{sect:meas}

In this section, we will describe our micromobility measurements campaign. We will start with the multi-stage methodology, and then will proceed with beam center micromobility emulation and real micromobility field measurements.

\subsection{Multi-Stage Methodology}

In absence of miniaturized equipment capable of operating in the sub-THz frequency band, we employ the multi-stage measurement methodology. At the first stage, we perform emulation of the UE beam center micromobility. To this end, by following \cite{stepanov2021statistical}, we utilize a laser pointer firmly attached to the UE running a certain application. A screen is utilized to track the motion of an imaginary beam center. At the second stage, we proceeded with field measurements using the sub-THz testbed operating at a 156 GHz frequency. Here, we utilized the beam center traces collected at the first stage to parameterize two goniometers with firmly connected transmitter (Tx) antenna. The goniometers allows to reproduce the movement of the UE's beam center traces exactly as were recorded at the first stage. Receiver (Rx) representing a BS was fixed and was not subjected to any movements. Below, we will proceed with detailed description of the experiments.

\subsection{Beam Center Micromobility}

\begin{figure}[b!]
\vspace{-0mm}
\centering
	\includegraphics[width=0.65\columnwidth]{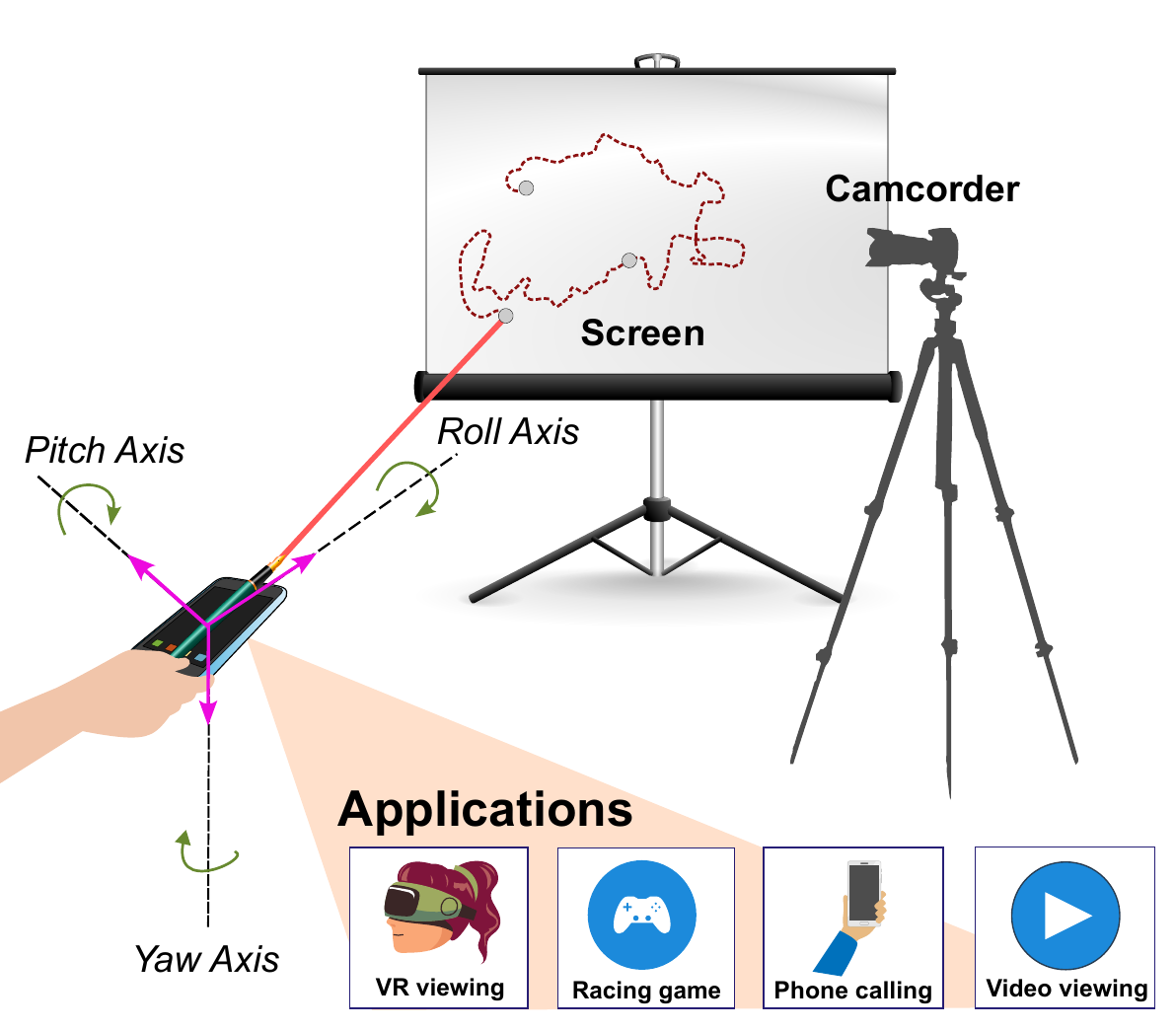}
\caption{The utilized measurement setup and environment.}
\label{fig:setup}
\vspace{-0mm}
\end{figure}

Fig. \ref{fig:setup} illustrates the experimental setup. We focused on four common smartphone applications: (i) video watching, (ii) VR viewing, (iii) voice calling, and (iv) gaming. While the first two applications are ``passive'' and do not control the actions of a user, racing games force specific user behavior, such as left and right movements, due to the use of a gyroscope for in-game vehicle control. VR applications can also disorient users in space.

A detailed analysis of the statistical characteristics of the beam center motion for the considered applications is presented in \cite{stepanov2021statistical}. Table \ref{tab:summary} summarizes the key findings. Even the seemingly similar applications of video watching and phone calling exhibit distinct statistical properties. VR viewing, also characterized by uncontrolled user movements, demonstrates the highest mobility speed. Notably, racing games exhibit a strong dependency in movement processes along the Ox and Oy axes, setting them apart from other applications. While no single characteristic can definitively distinguish these patterns and associate them with specific applications, a combination of factors, as employed in ML techniques, may enable precise classification.

% Modeling by Markov chains

After collecting the traces, we developed models for these time series. To accurately capture both distributional and correlational properties, we employed a two-dimensional Markov chain framework. We divided the entire range of Ox and Oy coordinates into an $N\times{}N$ grid. By calculating the transition probabilities between cells, we constructed the desired Markov model. The value of $N$ can be adjusted to achieve the desired model accuracy, with 100 being chosen based on the analysis in \cite{stepanov2021statistical}. As demonstrated in \cite{stepanov2021accuracy}, our two-dimensional Markov model outperforms simplified models such as decomposed Markov and Brownian motion models in terms of accuracy.

% Control traces

In real measurements of the received signal strength affected by the micromobility patterns we utilized high-precision goniometers allowing to control the direction of the antenna. We utilized the specified models to generate control traces consisting of 0s and 1s indicating positive and negative steps for vertical and horizontal planes. Note that the maximum time-resolution of the goniometers was $150$ ms/$^{\circ}$ which was slower than the micro-mobility speed of the racing game and VR watching applications. To this end, we utilized the time expansion technique allowing to align the maximum speed of the applications to that of the goniometers. The reverse procedure has been implemented to restore the actual time after measurements were carried out.

\subsection{Micromobility Field Measurements}

Here, we focus on the hardware implementation of a UE micromobility in ultra-directional sub-THz wireless channels. We provide details on the employed measuring equipment, including connection diagram and operating parameters, as well as the evaluated measurement accuracy.

\subsubsection{Sub-THz Wireless Modules}

\begin{figure}[!t]
\vspace{-0mm}
\centering
\includegraphics[width=0.75\columnwidth]{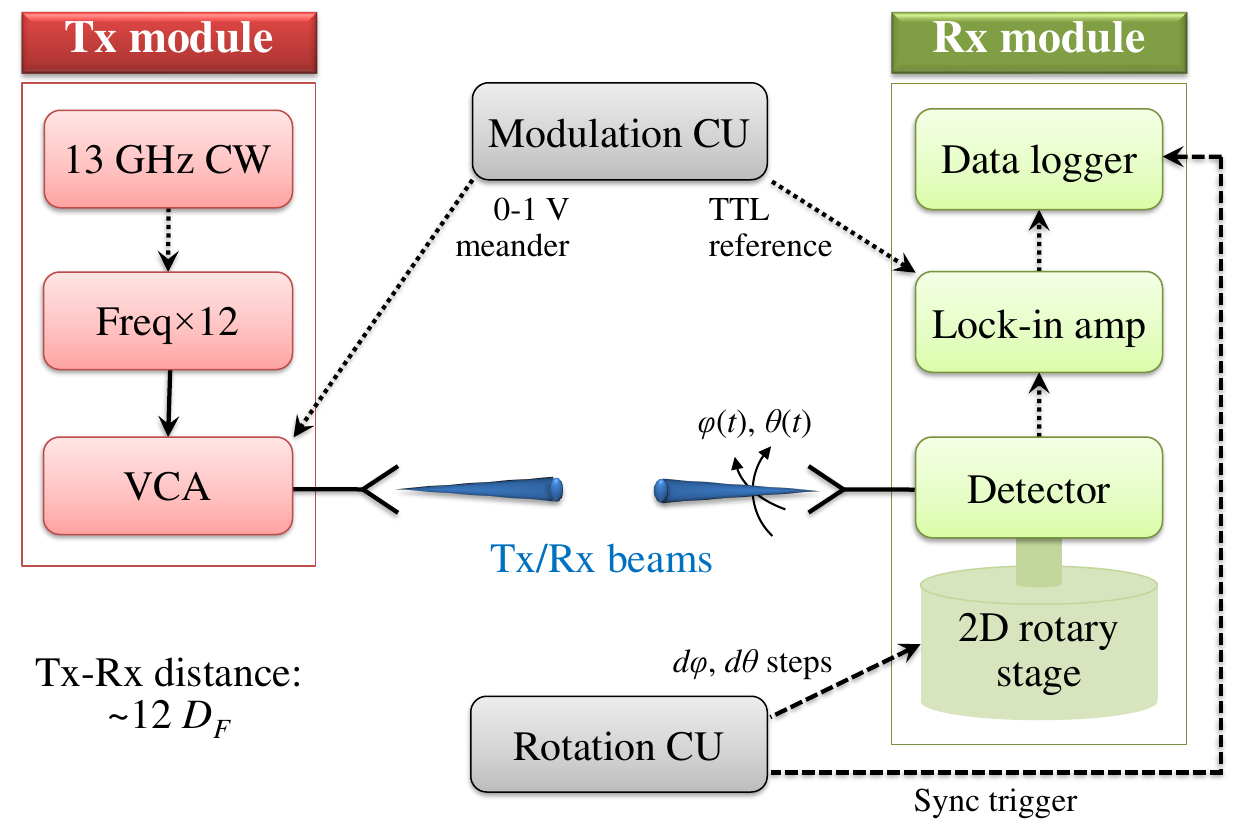}
\caption{Schematic diagram of the measurement setup.}
\label{fig:Setup-diagram}
\vspace{-0mm}
\end{figure}

Referring to Fig. \ref{fig:Setup-diagram}, the implemented experimental setup makes use of a static Tx and a mobile Rx modules. Tx is based on a sub-THz solid-state source utilizing a microwave constant waveform (CW) generator followed by a W-band frequency multiplication chain (Freq$\times$12), which is terminated by a voltage controllable attenuator (VCA). VCA is driven by a modulation control unit (CU) attached to its coax input, while its WR6.5 output is equipped with a narrow-beam pyramidal horn antenna. The assembly provides amplitude-modulated CW signals at carrier frequencies, $F_c$, of $132-162$ GHz with the corresponding modulation frequency of 25 kHz. The peak power available at the air interface equals 44 mW and is generated by Tx at $F_c$ = 156 GHz. Thus, we use this carrier frequency in all micromobility measurements.

In our measurements, UE micromobility is emulated in hardware using an Rx module with a steerable radiation pattern. Rx registers a Tx-generated signal incident on the aperture of the antenna-equipped Mott diode detector mounted on a two-dimensional (2D) rotary stage. The diode response voltage, $V_r$, is processed by a lock-in amplifier, whose readings are further logged. The data logger and the rotary stage are synchronized by a rotation CU such that the triplet of time-varying parameters including the received signal strength $P_r(t)$ and the pair of coordinate angles $\varphi(t), \theta(t)$ are acquired at each 2D step with fixed angular increments $d\varphi = d\theta = 1/145^\circ$. Here $P_r = 10 \, \log(V_r/S_u)+30$ with $S_u$ = 100 V/W denoting the diode detector responsivity at 156 GHz. 

\subsubsection{Motorized Mechanical Beam Steering}

\begin{figure}[!t]
\centering
\includegraphics[width=0.75\columnwidth]{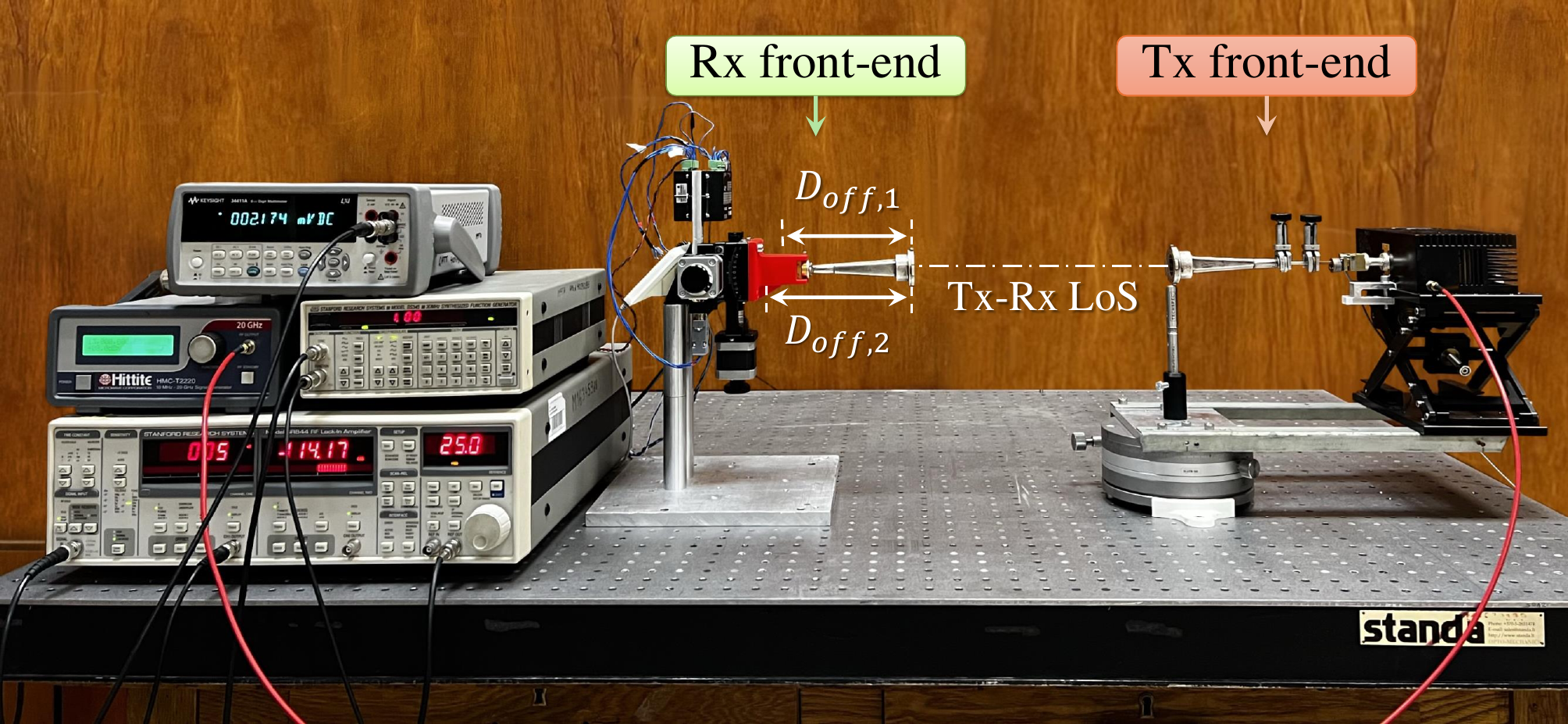}
\caption{On-the-bench photograph of the employed measuring equipment.}
\label{fig:Setup-photo}
\vspace{-3mm}
\end{figure}

\begin{figure}[!b]
\centering
\includegraphics[width=0.65\columnwidth]{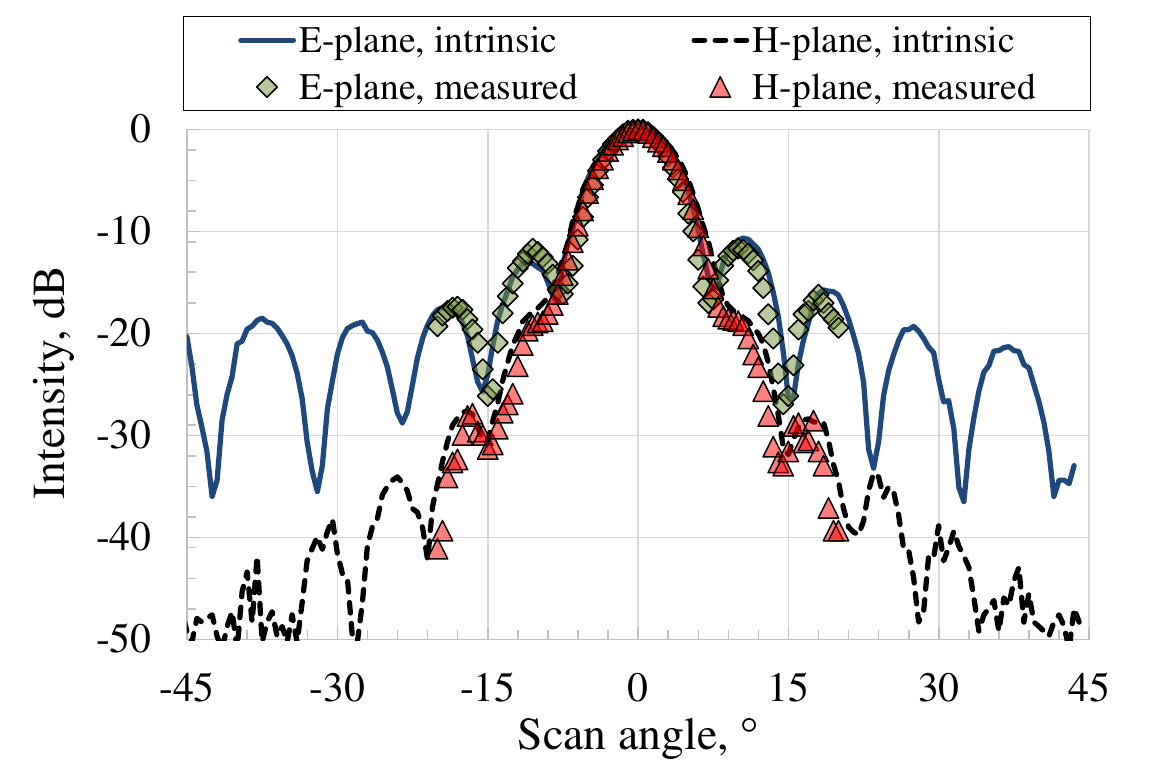}
\caption{Setup-dependent normalized Rx beams in the far-field zone.}
\label{fig:Beams-off}
\vspace{-0mm}
\end{figure}

%\textit{Motorized mechanical beam steering.} 
Fig. \ref{fig:Setup-photo} provides a photograph of the employed 2D rotary stage presented by a pair of cross-attached uniaxial motorized goniometers integrated with the antenna-equipped diode detector. This assembly enables Rx beam steering within $\pm20^{\circ}$ along both $\Vec{e}_\varphi$ and $\Vec{e}_\theta$ directions with a maximum scan speed of 6.7$^{\circ}$/s. However, it is characterized by offsets between the axes of the goniometers and the antenna aperture $D_{off,1} \approx D_{off,2} \approx 25$~cm. We evaluate impact of this feature on the far-field measurement accuracy by cross-comparing the intrinsic beam profiles of the employed horn antennas in H- and E-planes \cite{shurakov2023dynamic} with those measured by us in the current configuration of Rx beam steering. Fig. \ref{fig:Beams-off} provides results of the comparison. As one can clearly see, there are no notable differences between the expected and measured beam profiles. This can be explained by the fact that the far-field condition is reliably fulfilled, and a quasi-plane wavefront arriving at the Rx antenna is nearly constant in amplitude and phase when the antenna is rotated by $\pm20^\circ$ from its central position.  Indeed, the distance between Tx and Rx is fixed to 400 cm in all measurements while physical apertures of their antennas of 14 mm $\times$ 18~mm correspond to Fraunhofer distance $D_F$ = 33.7 cm for $F_c$~=~156~GHz.

\section{Statistical Characteristics}\label{sect:stat}

In this section, we report main statistical characteristics of the considered received signal strength and formalize the hypothesis for remote application detection.

\begin{figure*}[!t]
\vspace{-0mm}
\centering
\subfigure[{Racing game}]{
    \includegraphics[width=0.24\textwidth]{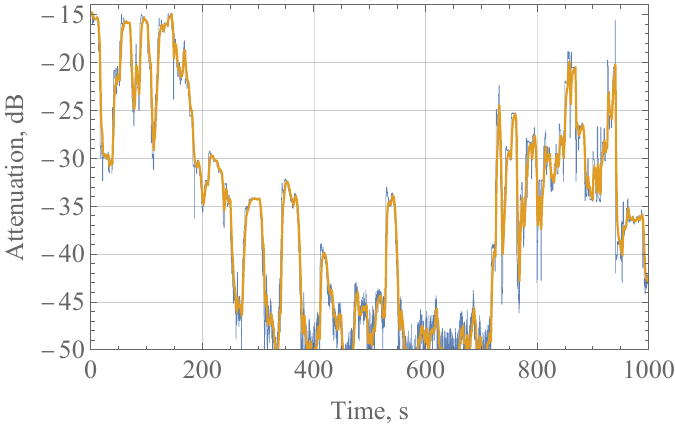}
    \label{fig:game_trace}
}~
\subfigure[{VR watching}]{
    \includegraphics[width=0.24\textwidth]{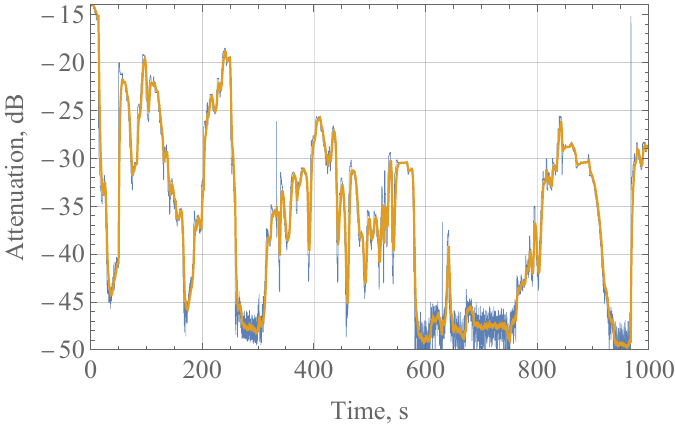}
    \label{fig:vr_trace}
}~
\subfigure[{Phone calling}]{
    \includegraphics[width=0.24\textwidth]{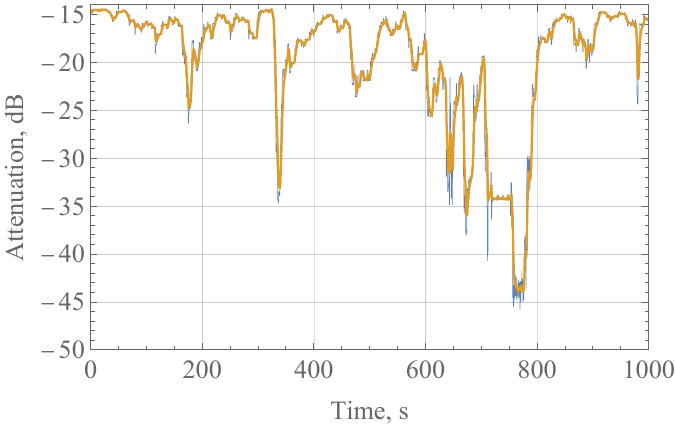}
    \label{fig:phone_trace}
}~
\subfigure[{Video watching}]{
    \includegraphics[width=0.24\textwidth]{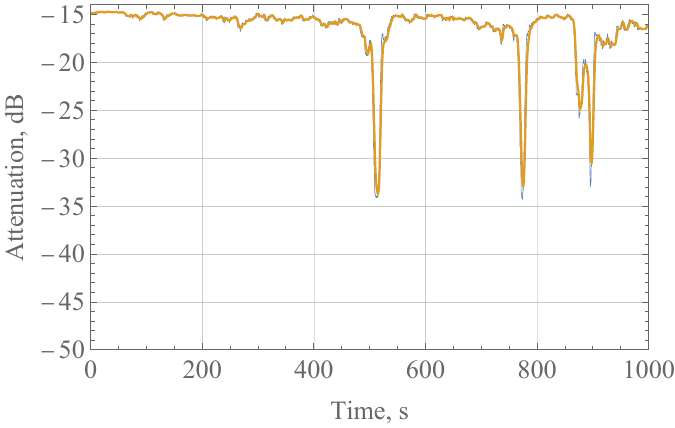}
    \label{fig:video_trace}
}
\caption{Time-series showing the sub-THz signal dynamics after beam tracking time instant for different applications.}
\label{fig:time-series}
\vspace{-0mm}
\end{figure*}

\subsection{Main Statistical Characteristics}

% Traces

We start discussing the results with sample traces of the received signal strength illustrated in Fig. \ref{fig:time-series} for all considered applications, where blue color shows the original data while yellow color highlighted exponentially smoothed weighting average (EWMA) with smoothing parameter $\gamma=0.001$. Although, we are mostly interested in the behavior of the considered metric right after the beam tracking time instant that happens at the time $t=0$, here, we specifically illustrate the long-term behavior of the time-series. 

%  Mian observations

Visual inspection of the presented time-series allows to identify a number of important observations. First of all, based on the received signal dynamics, one may differentiate between two classes of applications characterized by fast and slow micromobility. VR and racing game are characterized by fast drop in the received signal strength right after the beam tracking time instant and also show high variability over the whole duration of experiments. Although the reasons for this behavior are different, namely, explicit control by the player's actions in case of gaming and user disorientation for VR application, the ultimate effect is similar. Such applications naturally require high beam tracking frequency. Contrarily, video watching and phone calling applications show steady received signal strength behavior for long intervals after the beam tracking time instant. For such class of applications beam tracking can be seldom performed saving time and frequency resources of the system as well as UE power. %Finally, phone calling type of applications fall in between these two categories. Although the received signal strength deviates more noticeably as compared to video watching applications, the first notable drop is observed only at 170 s. 

\begin{figure}[!b]
\vspace{-0mm}
\centering
\subfigure[{Mean}]{
    \includegraphics[width=0.65\columnwidth]{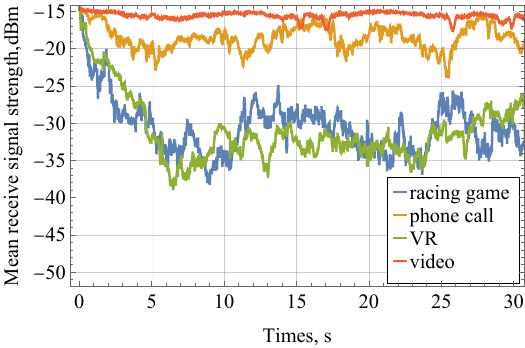}
    \label{mean}
}\\
\subfigure[{Standard deviation}]{
    \includegraphics[width=0.65\columnwidth]{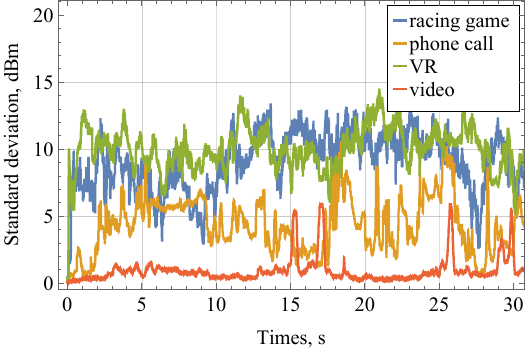}
    \label{dev}
}
\caption{Mean and standard deviation of the received signal strength averaged across all traces.}
\label{pic:mean_std}
\vspace{-0mm}
\end{figure}

\begin{figure}[!t]
\vspace{-0mm}
\centering
\subfigure[{Mean}]{
    \includegraphics[width=0.65\columnwidth]{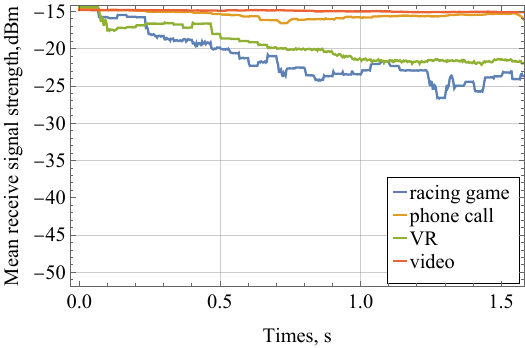}
    \label{fig:mean_zoomed}
}\\
\subfigure[{Standard deviation}]{
    \includegraphics[width=0.65\columnwidth]{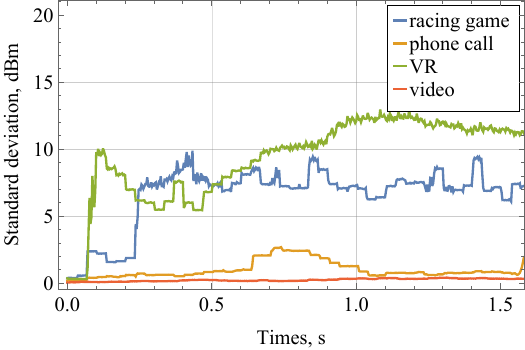}
    \label{fig:dev_zoomed}
}
\caption{Zoomed mean and standard deviation of the received signal strength just after beam tracking time instant averaged across all traces.}
\label{fig:mean_std_zoomed}
\vspace{-0mm}
\end{figure}

Having the main goal of differentiating between classes of applications based on their micromobility patterns right after beam alignment, we now proceed illustrating the mean and standard deviation of the time-series in Fig. \ref{pic:mean_std}. These illustrations have been obtained by ensemble averaging, that is, due to inherent non-stationarity of the time-series we calculated the mean value of all the micromobility traces for each application for a given time $t$.

By analyzing the presented results, we can indeed classify the considered applications into two categories (classes) based on the dynamics of the received signal strength, those having slow and fast micromobility behavior. VR watching and racing game applications are virtually indistinguishable based on the considered statistical parameters. For both applications, the ensemble average falls by 10 dB within 2 s, and by 20 dB within 5--6 s. Phone call falls in between almost completely stable video watching and highly dynamic VR watching and racing game applications. Specifically, the maximum degradation for video watching is just 2 dB over 30 s while for phone calling it decreases by at most 5 dB over the same interval.

Recall, that we are mostly interested in the behavior of the statistics just after the beam alignment time instant at $t=0$. Zoomed versions of the illustrations in Fig. \ref{pic:mean_std} for first 1.5 s are shown in Fig. \ref{fig:mean_std_zoomed}. Here, we can clearly observe different ``rates'' at which the mean and standard deviation values diverge from that at $t=0$. These rates can be utilized as a metric to discriminate different applications at the BS. Meanwhile, the values of standard deviation rise much quicker as compared to the fall of the mean values implying that utilizing standard deviation may potentially result in faster detection times.

% Drop to a level

\begin{table}[b!]
\vspace{-0mm}
\begin{center}\footnotesize
\caption{Minimum, mean, and maximum time for the received signal strength to fall below a certain threshold $S_{th}$ in ms}
\begin{tabular}{|l|l|l|l|l|}
 \hline
 \label{tab:threshold}
$S_{th}$, dB & Racing game & Phone calling & VR & Video\\
 \hline
3 &
\begin{tabular}{@{}l@{}}
6.9979\\70.2421\\246.693
\end{tabular} &
\begin{tabular}{@{}l@{}}
64.0063\\198.242\\385.04 
\end{tabular} &
\begin{tabular}{@{}l@{}}
7.334\\175.8177\\430.3766
\end{tabular} &
\begin{tabular}{@{}l@{}}
309.031\\N/A \\N/A
\end{tabular}\\
 \hline
5 &
\begin{tabular}{@{}l@{}}
6.9979\\ 87.2753\\ 274.694
\end{tabular} &
\begin{tabular}{@{}l@{}}
64.3397\\ 255.8403\\ 558.39 
\end{tabular} &
\begin{tabular}{@{}l@{}}
7.3341\\ 261.9927\\ 634.0633
\end{tabular} &
\begin{tabular}{@{}l@{}}
464.38\\ N/A \\ N/A
\end{tabular}\\
 \hline
7 &
\begin{tabular}{@{}l@{}}
6.9979\\ 92.0423\\ 278.0277
\end{tabular} &
\begin{tabular}{@{}l@{}}
70.6737\\ 471.38\\ 1865.8533 
\end{tabular} &
\begin{tabular}{@{}l@{}}
7.3341\\ 295.9297\\ 634.3967
\end{tabular} &
\begin{tabular}{@{}l@{}}
1506.4833\\ N/A \\ N/A
\end{tabular}\\
 \hline
10 &
\begin{tabular}{@{}l@{}}
23.6697\\ 96.0087\\ 281.3613
\end{tabular} &
\begin{tabular}{@{}l@{}}
208.354\\ 599.1333\\ 2182.55
\end{tabular} &
\begin{tabular}{@{}l@{}}
7.3340\\ 316.365\\ 734.407
\end{tabular} &
\begin{tabular}{@{}l@{}}
1510.8167\\ N/A \\ N/A
\end{tabular}\\
 \hline
15 &
\begin{tabular}{@{}l@{}}
24.3364\\ 114.8423\\ 291.0287
\end{tabular} &
\begin{tabular}{@{}l@{}}
211.688\\ 1013.0633\\ 2539.2533
\end{tabular} &
\begin{tabular}{@{}l@{}}
7.6674\\ 341.2\\ 751.0767
\end{tabular} &
\begin{tabular}{@{}l@{}}
1516.4833\\ N/A \\ N/A
\end{tabular}\\
  \hline
\end{tabular}
\end{center}
\end{table}

\begin{figure}[!b]
\vspace{-0mm}
\centering
\subfigure[{Mean}]{
    \includegraphics[width=0.65\columnwidth]{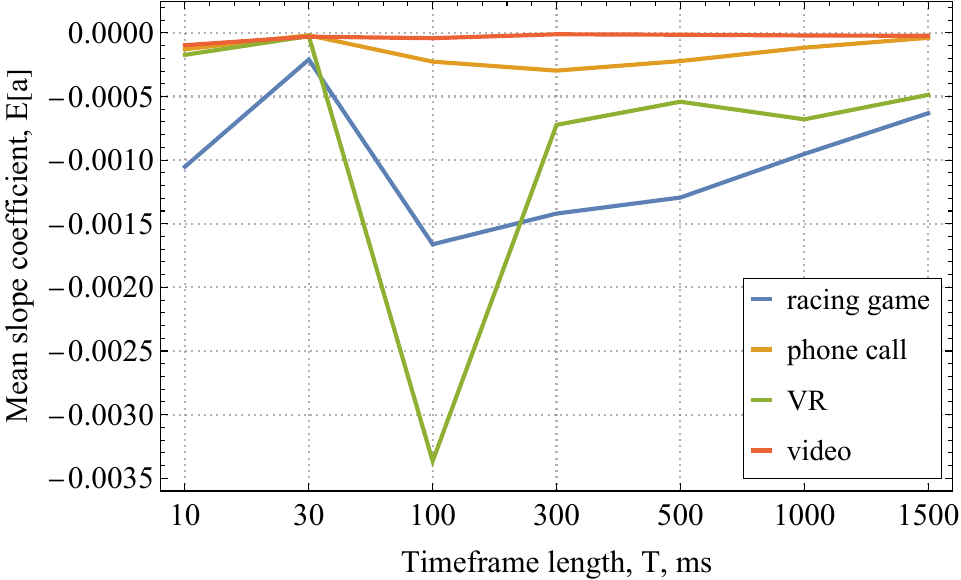}
    \label{fig:slope_mean}
}\\
\subfigure[{Standard deviation}]{
    \includegraphics[width=0.65\columnwidth]{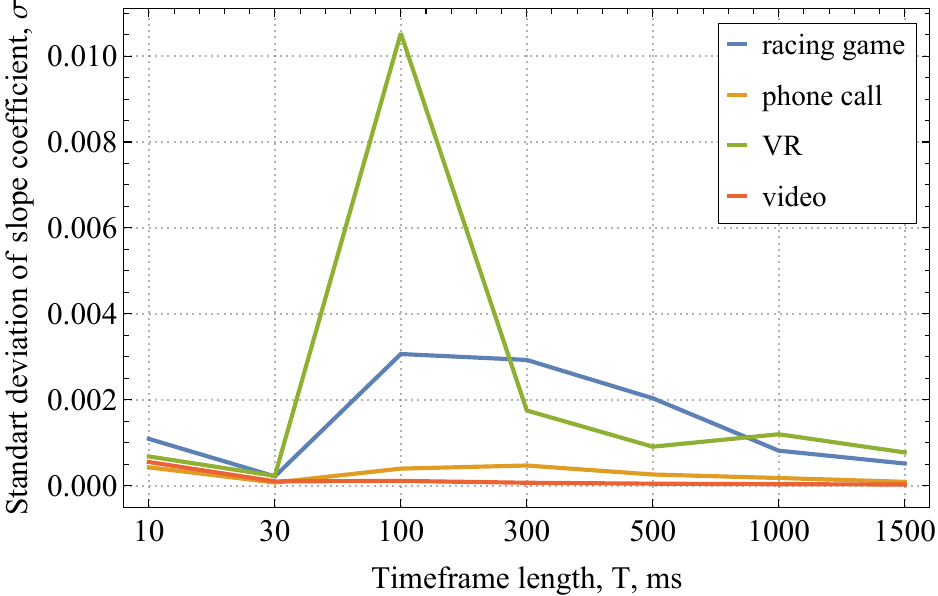}
    \label{fig:slope_std}
}
\caption{Mean and standard deviation of the regression slope as a function of the time window since the beam tracking time instant.}
\label{fig:slope}
\vspace{-0mm}
\end{figure}

\subsection{Time to Fall Below a Given Level}

Modern cellular systems including 5G NR utilizing the directional antennas perform beam searching procedure at regular time instants. Even for such systems, the amount of overhead associated with it can reach 25\% of all the resources available \cite{nrPhy}. When the number of possible antenna configurations will increase in 6G sub-THz systems due to even larger antenna arrays there might be the need for on-demand beam searching. It has been shown in \cite{petrov2020capacity} that this approach can outperform regular beam tracking for antenna arrays larger than $128\times{}128$ elements. Thus, the time it takes for the received signal strength to fall below a given level is of interest.

Table \ref{tab:threshold} illustrates the minimal, mean, and maximal time it takes for the received signal strength to fall by 3, 5, 7, 10, and 15 dB with respect to the level at time $t=0$ when antennas are perfectly aligned, where we utilize ``N/A'' when the considered level is not achieved. Note that this time is averaged over all experiments associated with a given application. By analyzing the presented results, we observe that racing game is characterized by the quickest mean time to fall below all the considered threshold. However, already for VR application the mean time is at least twice bigger for almost all thresholds. For small values of the threshold, i.e., 3 dB and 5 dB, phone call is characterized by mean times similar to VR watching applications while for bigger values it is at least twice bigger. Finally, for video watching application, there were no experiments, where the received signal strength fell below by even 3 dB.

\subsection{Main Hypothesis}

As discussed previously, the considered applications are characterized by different behavior after the beam tracking time instant. Thus, one of the metrics, that can be utilized to discriminate application is the slope of the received signal strength. To illustrate it visually, Fig. \ref{fig:slope} shows the the mean and standard deviation of the regression slope coefficient estimated by utilizing the raw (non-smoothed) received signal strength over different intervals in the form $(0,T)$ measured from the beam tracking time instant, $t=0$ for different values of $T$. To produce these data, we utilized the least-squares fit method (LSF) over ensemble averaged data.

%\textcolor{red}{Here goes main illustrations of the slope behavior. We need: mean and standard deviation of the slope for different time time instants from the beam alignment point, 20, 30, 100, 200, 500.}  As one may observe, the coefficients for low and high micromobility applications...

% Analysis

By analyzing the data presented in Fig. \ref{fig:mean_std_zoomed}, one may observe that the maximum difference is attained at the time interval $(0-100)$ ms implying that the slope over these interval may serve as a factor to distinguish between different classes of applications. However, as one may observe, the standard deviation corresponding to this maximum difference already increases. This may imply that classic statistical tests detecting the difference between mean values may not provide sufficient accuracy, thus, requiring more powerful approaches such as those available in the class of ML algorithms.

% PCA

\begin{figure}[!t]
\vspace{-0mm}
\centering
\subfigure[{0 - 100 ms window}]{
    \includegraphics[width=0.65\columnwidth]{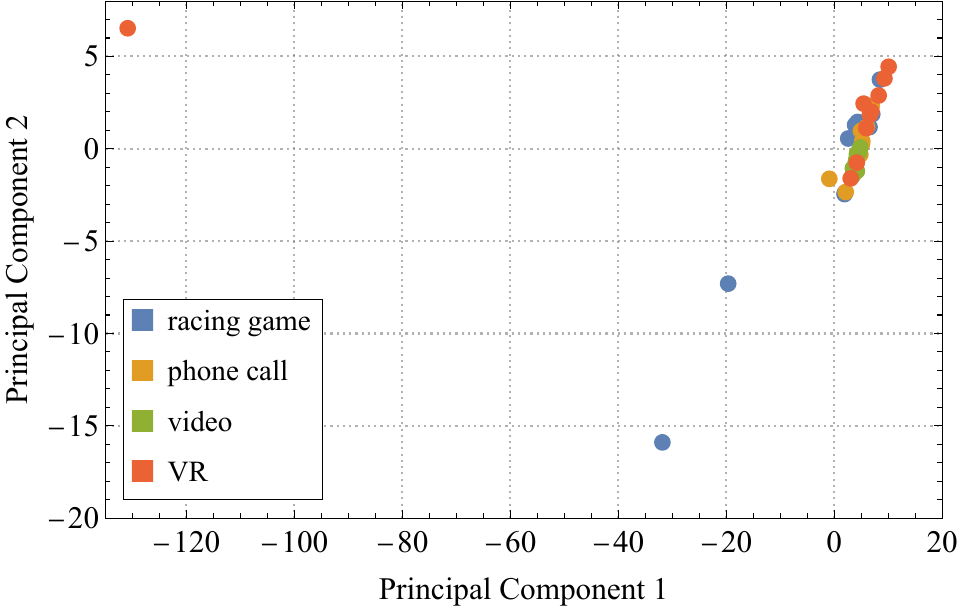}
    \label{mean}
}\\
\subfigure[{50 - 150 ms window}]{
    \includegraphics[width=0.63\columnwidth]{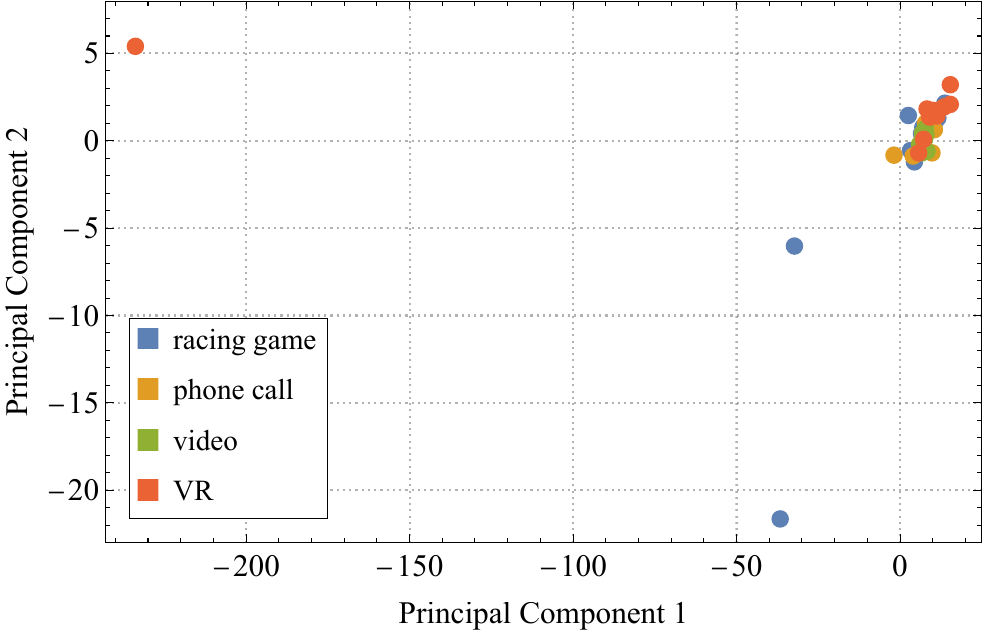}
    \label{dev}
}
\caption{The principal component analysis for all traces.}
\label{pic:mean_std_pca}
\vspace{-0mm}
\end{figure}

To statistically assess whether the data are separable, we performed the principal component analysis (PCA, \cite{greenacre2022principal}) for two statistical data windows, $(0-100)$ ms and $(50-150)$ ms. Recall, that generally PCA allows to reduce the dimensions of data producing new variables that contain most of the information about the original data set. The results for the first two components are presented in Fig.~\ref{pic:mean_std_pca}. By analyzing the data, we observe that these components do not produce perfect linear combination that separates different clusters corresponding to micromobility of different applications. However, we still see that racing game and VR applications have some values that are drastically different from video and phone call applications implying that there should be statistical methods to distinguish between applications with slow and fast micromobility.

\section{Real-Time Applications Classification}\label{sect:class}

In this section, we begin with specifying the overall applications class detection procedure based on the received signal strength. Then, we proceed introducing two types of tests for detecting the application class.

\subsection{The Overall Procedure}

% The task

The task of the algorithm designed in this section is to determine the type of application or at least its class. By utilizing this information, BS then may estimate the optimal interval between beam tracking time instants. This is done by utilizing the pessimistic assessment of the signal drop time to a given threshold reported in Table \ref{tab:threshold}. Alternatively, to avoid outages caused by beam tracking time interval being too large, one may utilize $x$-quantile of the signal drop time cumulative distribution function (CDF), where $x$ is set to a values close to unity, e.g., 0.95.

% Assumption

We assume that reliable statistics of slopes for different applications is available. It can be obtained by performing experiments similar to those described in Section \ref{sect:meas}. Statistically speaking, these data serve as initial population.

% The overall procedure description

The overall operational cycle is logically divided into two phases, see Algorithm \ref{alg:1}: (i) warm-up phase and (ii) active phase. Initially, at the warm-up phase, BS does not have any information regarding the type of application utilized at UE. Thus, it instructs UE to utilize the minimum (default) interval between SSB synchronization time instants. Once the application is activated, that is, UE initiated protocol data unit (PDU) session, UE starts to collect the received signal strength data. As a single realization of the process is not sufficient, UE keeps utilizing the same beam tracking interval. The process is repeated multiple times to gain reliable statistical data. 

During the beam tracking interval BS records statistical data and then applies classification procedure by utilizing the accumulated data and those available from the population assessing whether the estimated data belongs to one of the known application or application classes. Once the classifier returns a positive decision with respect to one of the available applications' or application classes, BS estimates the beam tracking interval and instructs UE to utilize it. This signals the start of the active phase, where the optimal beam tracking interval can be utilized minimizing the overheads associated with the beam tracking procedure.

\begin{algorithm}[t!]
    \caption{Beam Tracking Interval Detection}\label{alg:1}
    \KwIn{$N$ -- intervals between beam tracking, $R$ - an array containing signal strength values}
    \KwOut{Optimal beam tracking interval}
    Phase 1: Warm-up. Initialize the beam tracking interval to the minimum (default) interval\\
    %between SSB synchronization time instants\\
    \For{$i=1,2,\ldots,N$}{
        Measure received signal strength as $r_i$\\
        Append $r_i$ to $R$
    }
    Phase 2. Active. Calculate classification features: slope, mean, variance, etc. from the array $R$\\
    Apply application/class test\\
    \eIf{features indicate suboptimal beam tracking interval}{Increase beam tracking interval}{Maintain the beam tracking interval}
    Apply the new beam tracking interval
\end{algorithm}

% What do later

Despite the application is conventionally utilized for longer periods of time as compared to the beam tracking interval, user may eventually change the application. This may result in outages happening earlier than the currently utilized beam tracking interval (e.g., when slow mobility application is replaced by the fast mobility one) and may also result in inefficient use of resources, when user decides to start using slow mobility application instead of high-mobility one. Thus, when the application is detected BS continues to monitor statistical data between beam tracking time intervals. The system turns back into the warm-up phase if one of the following events is detected: (i) the link is in outage condition prior to the expiration of the beam tracking interval or (ii) tests assessing the identified classification start to fail.

% Backing up the choice of tests

\subsection{Conventional Statistical Approach}

A critical part of the Algorithm \ref{alg:1} is detection of the application or its class, based on the statistical data gathered during the warm-up phase. To decrease the burden in terms of the computational efforts, we first look for a solution in the class of conventional statistical tests and the statistics of interest are the slopes of received signal strength obtained during the warm-up phase.

We propose to classify applications or application classes based on the non-parametric statistical tests as the distribution of the population is not known apriori. The chosen test is Mann-Whitney $U$ test based on ranking statistics \cite{papoulis1990probability,rosner1999use} as it is one of the most reliable tests for small samples, where the number of data points (beam tracking intervals measured over the warm-up phase) is less than $30$. This tests assess the null hypothesis $H_0$ consisting of that, for randomly selected values $X$ and $Y$ from two populations, the probability of $X$ being greater than $Y$ is equal to the probability of $Y$ being greater than $X$. That is, $H_0$ implies that two samples come from the same population, while the alternative hypothesis $H_1$ implies that $X$ and $Y$ come from different populations and thus differ.

\subsection{Machine-Learning Classification}

The data presented in Fig. \ref{fig:slope} show that not only the mean values of slow and fast micromobility applications differ but standard deviation too. This may lead to inability of the conventional statistical test to differentiate between applications. To this end, we also propose to utilize ML-based algorithms that are well-suited for classification tasks. Accounting for the fact that these algorithms need to learn on-line the amount of statistical data is limited. Thus, we omit the use of ``heavy-weight'' neural network classifiers that require large amount of data for training purposes and limit our attention to feature-based classifiers including trees, random forests.

\subsubsection{Tree Classifiers}

The first class of algorithms we will consider is a decision tree. A decision tree is a powerful data analysis tool that recursively partitions a dataset into smaller, more manageable subsets. By prioritizing attributes that maximize information gain, decision trees minimize overall uncertainty within the data. This involves evaluating the reduction in entropy achieved by splitting the data based on different attributes.

Recall that the entropy of a dataset is computed as
\begin{align}
H(D) = - \sum (p_i \log_2(p_i)),
\end{align}
where $p_i$ is the frequency of class $i$ within the dataset.

The information gain when considering a feature $F$ in a dataset $D$ is given by
\begin{align}
I(D,F) = H(D) - H(D|F),
\end{align}
where $H(D|F)$ is the conditional entropy left after considering the feature $F$.

Decision trees recursively divide datasets based on attributes that offer the greatest information gain, continuing until predefined criteria, such as a specified tree depth or minimum samples per node, are met. This structured approach simplifies data interpretation and enhances its predictive capabilities.

\subsubsection{Random Forest Classifiers}

Random forest is an ensemble machine learning method that enhances the accuracy of decision trees. By creating a forest of multiple trees, Random forest increases robustness and mitigates the risk of overfitting. The algorithm begins by generating bootstrap samples, which are subsets of the original data created through random sampling with replacement. A decision tree is then built for each bootstrap sample. To foster diversity among the trees, Random forest randomly selects a subset of features for node splitting at each level of the tree-building process. Unlike a single decision tree, the trees in a Random forest are typically grown to their full size without pruning.

Random forest traditionally relies on majority voting for classification tasks. However, it can also determine the final decision based on the median of outcomes from all decision trees, providing a robust measure that minimizes the influence of outliers. This median is computed as 
\begin{align}
\hat{Y}=M(Y_1,Y_2,\dots,Y_t),
\end{align}
where $Y_1, Y_2, \dots,Y_t$ are the decisions from the individual trees.

Random forest effectively handles outliers and noisy data, providing valuable insights into the features that drive predictions. Its use of the median further strengthens its resilience to data anomalies. While its ensemble nature increases computational complexity and may reduce interpretability, Random forest remains a popular choice for its accuracy across diverse data types and machine learning tasks.

\subsubsection{Considered Features}

Both trees and random forests require attributes (features) to be used for classification. We utilize the following. The first chosen attribute coincides with the statistics utilized for conventional test-based classification, that it, the slope of the regression obtained using the LSF for partially overlapping $(0,T)$ intervals, $T=10,20,\dots,300$ ms. In addition, we also utilize the direct statistics, related to the mean, variance, lag-1 autocorrelation for $(0,T)$ intervals. Since the data variability for different applications is evident in Figs. \ref{pic:mean_std} and \ref{fig:mean_std_zoomed}, we also utilize an attribute related to the spectral domain data representation \cite{zhinuk2024spectral}. As a special attribute we also use the summed periodogram statistics over these intervals obtained via short-term Fourier transform (STFT) defined as
\begin{align}\label{fourier_magic}
S(m,\omega) = \sum_{n=0}^{W} f[m+n] w[n] e^{-j \omega n},
\end{align}
where $S(m,\omega)$ represents the STFT magnitude at time $m$ and the frequency $\omega$, $f[m+n]$ is the $(m+n)$-th sample of in the time window $(0,T)$, $w[n]$ is the $n$-th sample of the window function. Note that in order to capture spectral domain variations, instead of STFT, one may utilize any transform that converts a signal from the time to the frequency domain.

%The second chosen attribute is the mean value of non-overlapping intervals, $(T_1,T_2)$, $T_2=10,20,\dots,300$ ms, $T_2-T_1=10$ ms. 

Once the set of attributes for each trace was calculated, the next step was to use them for classification. For this purpose, Python was used, specifically its built-in implementations of considered classifiers, see Section \ref{sect:numerical} for details. To classify the data, the attributes have been provided to the utilized classifiers, where each feature was marked with the scenario of use to which it belonged. %As a training sample, we utilize 10 time-series for each applications resulting in $300$ data points for all the considered attributes for each application. As a testing sample, we utilized 20 separately measured time-series.

\subsubsection{Metrics of Interest}

To evaluate the performance of the proposed algorithms we utilize two metrics of interest: (i) accuracy and (ii) F1. The former, $A$ is a basic metric for evaluating the quality of a classification. Specifically, it measures the proportion of correctly classified examples out of the entire dataset. This metric is useful when the classes are balanced, meaning that the number of examples in each class is approximately the same, which is the case in our datasets. The metric is calculated as
\begin{align}
A = \frac{N^{\star}}{N}.
\end{align}
where $N^{\star}$ is the number of correctly classified examples, $N$ is the total number of examples.

The F1 score combines precision and recall into a single measure by calculating their harmonic mean. This balances the impact of both precision and recall on the model's quality assessment, making the F1 score especially useful when both performance aspects need consideration.

In our study, we first calculate the F1 score separately for each class, then average the obtained F1 scores. This approach considers the contribution of each class equally, regardless of class size. The following have been used to calculate F1 score
\begin{align}
F1 = \frac{1}{2} \sum_{i=1}^3 F1_i,
\end{align}
where individual scores are given by
\begin{align}
F1_i = 2 \frac{P_i R_i}{P_i + R_i}, \, i = 1, 2,
\end{align}
$P_i$ and $R_i$ are precision and recall estimated as
\begin{align}
P_i = \frac{TP_i}{TP_i + FP_i}, \,
R_i = \frac{TP_i}{TP_i + FN_i}, \, i = 1, 2,
\end{align}
where $TP_i$ are the true positive cases for class $i$, $FP_i$ are the false positive cases for class $i$, i.e., cases where other classes were mistakenly classified as class $i$, and $FN_i$ are the false negative cases for class $i$, i.e., cases where class $i$ was mistakenly classified as another class. %When classification of individual applications was performed, the number of classes were set to 4.

\section{Numerical Results}\label{sect:numerical}

In this section, we report out numerical results related to classification of applications based on received signal strength data. We first start with conventional statistical approach and then proceed reporting the data for tree and forest classifiers. We report classification results for slow and fast micromobility applications as well as for fine-grained classification attempting to differentiate between individual applications.

\subsection{Mann-Whitney Classification}

% Individual apps

We begin with the results of utilizing the Algorithm \ref{alg:1} based on the Mann-Whitney test for classification for slow/fast micromobility applications and for individual applications based on the slope behavior. To this end, Table \ref{tab:allAppsMW} shows $p$-values of the pairwise comparison between slopes of different applications for confidence probability $0.95$ over two time periods, $0-100$ ms and $50-150$ ms, for 10 and 30 experiments corresponding to beam tracking intervals. Recall that one accepts the null hypothesis $H_0$ consisting in that samples belong to different populations when the $p$-value is smaller than 0.05 for $0.95$ confidence probability. By analyzing the presented data one may observe that for both values of the time periods and even for 30 time series one cannot reliably distinguish between individual applications. The rationale is that, as we observed in Fig. \ref{fig:slope}, not only the means of slopes are different for different applications bit variance too. The latter affects the accuracy of the Mann-Whitney test.

\begin{table}[t!]
\vspace{-0mm}
\begin{center}
\caption{$p$-value of the Mann-Whitney test for individual applications}
\begin{tabular}{|l|l|l|l|l|}
 \hline
\multicolumn{5}{|l|}{\textbf{Slopes computed over 0-100 ms, 10 time-series}} \\ \hline\hline
Application & Racing & Calling & Video & VR\\
 \hline\hline
Racing & 0.967 & 0.344 & 0.045 & 0.140 \\ \hline
Call   & 0.344 & 0.913 & 0.140 & 0.161 \\ \hline
Video  & 0.045 & 0.140 & 0.982 & 0.791 \\ \hline
VR     & 0.140 & 0.161 & 0.791 & 0.892 \\ \hline
\multicolumn{5}{|l|}{\textbf{Slopes computed over 50-150 ms, 10 time-series}} \\ \hline\hline
Application & Racing & Calling & Video & VR\\
 \hline\hline
Racing & 0.945 & 0.212 & 0.014 & 0.185 \\ \hline
Call   & 0.212 & 0.921 & 0.140 & 0.850 \\ \hline
Video  & 0.014 & 0.140 & 0.974 & 0.088 \\ \hline
VR     & 0.185 & 0.850 & 0.088 & 0.901 \\ \hline
\multicolumn{5}{|l|}{\textbf{Slopes computed over 0-100 ms, 30 time-series}} \\ \hline\hline
Application & Racing & Calling & Video & VR\\
 \hline\hline
Racing & 0.967 & 0.344 & 0.045 & 0.140 \\ \hline
Call   & 0.344 & 0.911 & 0.140 & 0.161 \\ \hline
Video  & 0.045 & 0.140 & 0.951 & 0.791 \\ \hline
VR     & 0.140 & 0.161 & 0.791 & 0.949 \\ \hline
\multicolumn{5}{|l|}{\textbf{Slopes computed over 50-150 ms, 30 time-series}} \\ \hline\hline
Application & Racing & Calling & Video & VR\\
 \hline\hline
Racing & 0.972 & 0.212 & 0.014 & 0.185 \\ \hline
Call   & 0.212 & 0.916 & 0.140 & 0.850 \\ \hline
Video  & 0.014 & 0.140 & 0.947 & 0.088 \\ \hline
VR     & 0.185 & 0.850 & 0.088 & 0.967 \\ \hline
\end{tabular}
\label{tab:allAppsMW}
\end{center}
\end{table}

\begin{table}[t]
\vspace{-0mm}
\begin{center}\small
\caption{$p$-value of the Mann-Whitney test for application's classes}
\begin{tabular}{|l|l|l|}
\hline
\multicolumn{3}{|l|}{\textbf{Slopes computed over 0-100 ms, 10 time-series}} \\ \hline\hline
Application type & Slow micromobility & Fast micromobility\\
\hline\hline
Fast micromobility & 0.856 & 0.155 \\ \hline
Slow micromobility & 0.155 & 0.793 \\ \hline
\multicolumn{3}{|l|}{\textbf{Slopes computed over 50-150 ms, 10 time-series}} \\ \hline\hline
Application type & Slow micromobility & Fast micromobility\\
\hline\hline
Fast micromobility & 0.878 & 0.041 \\ \hline
Slow micromobility & 0.041 & 0.823 \\ \hline
\multicolumn{3}{|l|}{\textbf{Slopes computed over 0-100 ms, 30 time-series}} \\ \hline\hline
Application type & Slow micromobility & Fast micromobility\\
\hline\hline
Fast micromobility & 0.912 & 0.242 \\ \hline
Slow micromobility & 0.242 & 0.834 \\ \hline
\multicolumn{3}{|l|}{\textbf{Slopes computed over 50-150 ms, 30 time-series}} \\ \hline\hline
Application type & Slow micromobility & Fast micromobility\\
\hline\hline
Fast micromobility & 0.937 & 0.013 \\ \hline
Slow micromobility & 0.013 & 0.893 \\ \hline
\end{tabular}
\label{tab:classAppsMW}
\end{center}
\end{table}

% Classes

Consider now, whether the conventional methods can be used to differentiate between application's classes. Table \ref{tab:classAppsMW} shows $p$-values of the pairwise comparison between different slopes of classes of applications computed for confidence probability $0.95$ over two time periods, $0-100$ ms and $50-150$ ms, for 10 and 30 experiments. As one may observe, the utilized test allows to statistically differentiate between classes of applications already for 10 experiments when the time-interval $(50-100)$ ms is considered. For higher number of experiments, 30, the $p$-values becomes even less. For the time interval $(0,100)$ ms, the alternative hypothesis needs to be taken in both cases although the $p$-value is much smaller for 30 experiments than for 10 experiments. The explanation of this fact can be deduced from the raw traces illustrated previously in Fig. \ref{fig:time-series}. Indeed, for a certain rather small amount of time after the beam tracking time instant (happening at $t=0$), received signal strength remains almost intact. The reason is the way the experiments were conducted restarting the application for each new measurement. Note that in real situations this behavior is not typical and thus one may expect reliable detection over the interval $(0,100)$ ms.

% Conclusions

We note that according to the algorithm in Section \ref{sect:class} reliable detection is feasible after more than 10 beam tracking time instants. By utilizing the maximum feasible interval defined for 5G NR, 320 ms, it implies that the delay associated with application detection is on the order of few seconds. To improve the performance further we now proceed considering ML-based classification performance.

\subsection{ML-based Classification}

\begin{table}[b]
\vspace{-0mm}
\begin{center}\scriptsize
\caption{Mean values of features selected for ML-based classification}
\begin{tabular}{|l|l|l|l|l|}
\hline
Features & Racing & Call & Video & VR \\ \hline\hline
Slope        & -0.000712     & -0.0000984 & -0.0000966 & -0.00036313   \\ \hline
Mean         & -14.761       & -14.796 & -14.846 & -14.758   \\ \hline
Variance     & 0.00358       & 0.00027 & 0.00017 & 0.00101   \\ \hline
100 ms value & -14.776       & -14.801 & -14.849 & -14.773   \\ \hline
STFT sum     & 21794.4       & 21895.2 & 22042.1 & 21788.8  \\ \hline
Lag-1  & 0.61514       & 0.60434 & 0.78021 & 0.6403  \\ \hline 
\end{tabular}
\label{tab:features_app}
\end{center}
\end{table}

\begin{table}[t]
\vspace{-0mm}
\begin{center}\small		
\caption{Mean values of features selected for ML mobility class classification}
\begin{tabular}{|l|l|l|}
\hline
Features & Fast application class & Slow application class \\ \hline\hline
Slope        & -0.0002       & -0.0004   \\ \hline
Mean         & -14.802       & -14.778   \\ \hline
Variance     & 0.00059       & 0.00193   \\ \hline
100 ms value & -14.811       & -14.788   \\ \hline
STFT sum     & 21915.4       & 21844.8   \\ \hline
Lag-1  & 0.71026       & 0.60974   \\ \hline             
\end{tabular}
\label{tab:features_class}
\end{center}
\end{table}

\begin{figure*}[t!]
	\centering
	\subfigure[{Accuracy}]{
        \includegraphics[width=0.32\textwidth]{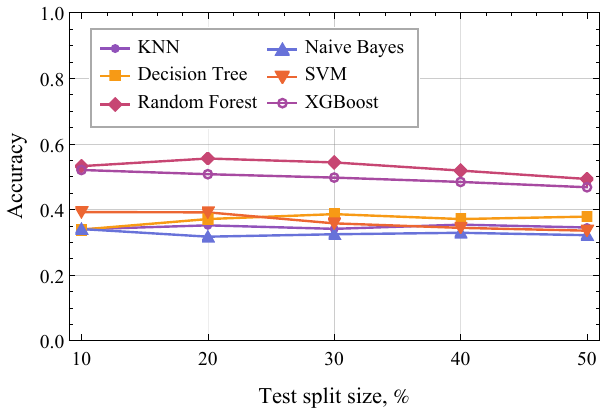}
		\label{fig:app_accuracy}
	}~
    \subfigure[{Recall}]
	{
		\includegraphics[width=0.32\textwidth]{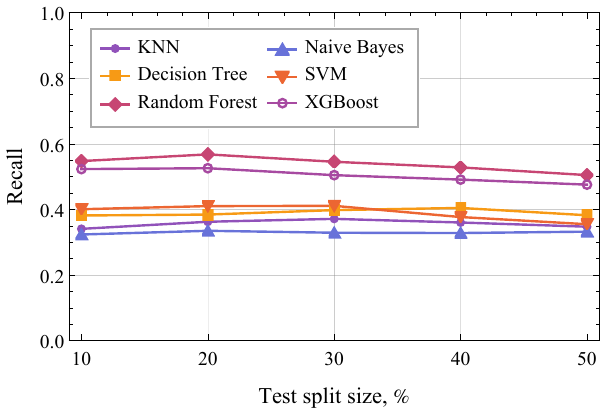}
		\label{fig:app_recall}
	}~
	\subfigure[{F1 score}]
	{
        \includegraphics[width=0.32\textwidth]{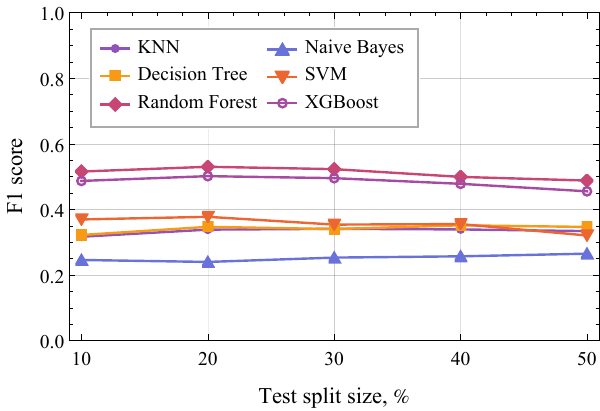}
		\label{fig:app_f1}
	}
	\caption{Performance metrics for mobility application classification.}
	\label{fig:application}
\end{figure*}

We now proceed assessing whether ML-based classification can outperform conventional Mann-Whitney statistical test. To this end, in addition to the decision tree and random forest described in Section \ref{sect:class}, we also considered even simpler algorithms such as K nearest neighbours (KNN) and Naive Bayes. To assess the upper bound of tree-based classifiers we consider one of the most advanced algorithms -- XGBoost \cite{chen2016xgboost}. Finally, we also added Support Vector Machine (SVM) as one of the algoritms that does not require very large training sets.

Our code is implemented in Python, utilizing KNeighborsClassifier, DecisionTreeClassifier, RandomForestClassifier, GaussianNB, SVM, and XGBClassifier libraries. The dataset used for training the models consisted of features derived from all the available received signal strength time-series (120 overall). These features include: (i) slope coefficient, (ii) mean, (iii) variance, (iv) 100 ms signal strength value, (v) STFT sum, and (vi) lag-1 autocorrelation. The mean values of these features are presented in Tables \ref{tab:features_app} and \ref{tab:features_class} for inidvidual applications and for application classes. As shown in the tables, classification was performed in two ways: by specific application (racing, call, video and VR) and by application class (e.g., fast mobility vs. slow mobility). Three evaluation metrics were considered: accuracy, recall, and F1-score.

Due to the limited amount of data, moderate tuning of the algorithms was used to prevent overfitting. The KNN model utilized 3 nearest neighbors; the Decision Tree was restricted to a depth of 3; and the random forest model was built with 100 trees. For tuning the SVM, we employed the GridSearchCV function to optimize three hyperparameters: C, kernel, and gamma. The grid search recommended using the default Radial basis function (RBF) kernel and a default value of $C = 1$, with only the gamma value adjusted to $0.01$. The gamma parameter determines the extent of influence that a single training example has; a small value of gamma indicates that the influence of individual examples is broad. This choice is justified by the simplicity of the constructed model, which does not require capturing complex patterns or shapes in the data -- a reasonable approach given the small size of the dataset. We also performed a GridSearchCV to tune the hyperparameters of the XGBClassifier, including eta, gamma, max\_depth, and min\_child\_weight. All parameters were set to their default values except for $\eta$. This parameters, which represents the learning rate, was set to $0.15$, which facilitates finding a more optimal solution but increases the computational complexity. 

\begin{figure*}[t]
	\centering
	\subfigure[{Accuracy}]{
        \includegraphics[width=0.32\textwidth]{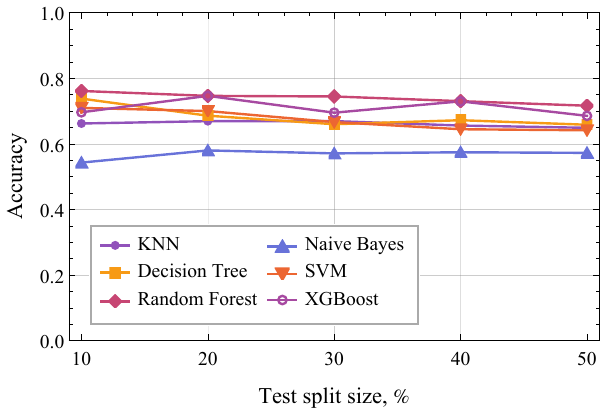}
		\label{fig:class_accuracy}
	}~
    \subfigure[{Recall}]
	{
		\includegraphics[width=0.32\textwidth]{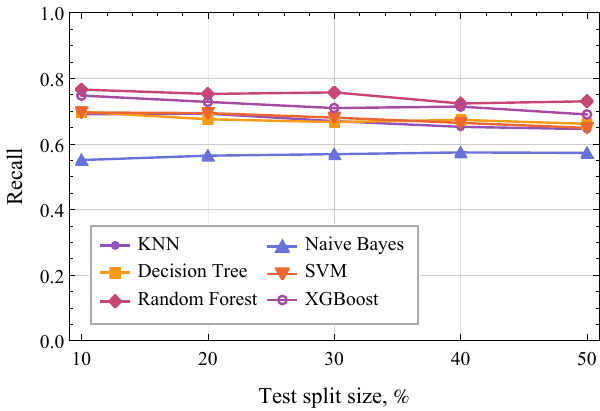}
		\label{fig:class_recall}
	}~
	\subfigure[{F1 score}]
	{
        \includegraphics[width=0.32\textwidth]{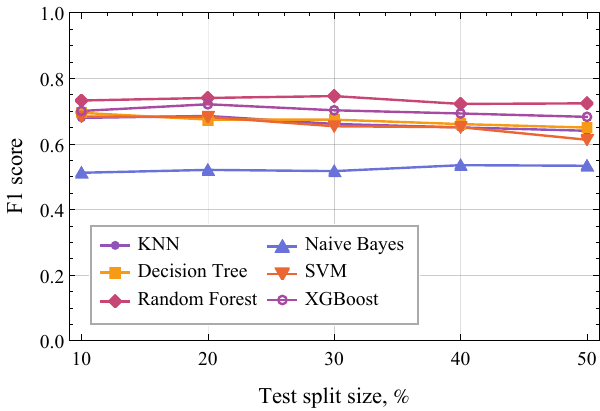}
		\label{fig:class_f1}
	}
	\caption{Performance metrics for mobility class classification.}
	\label{fig:class}
\end{figure*}

We begin with individual application classification. As shown in Fig.~\ref{fig:application}, the random forest algorithm achieves the best results across all three metrics. However, even this algorithm does not achieve an accuracy higher than $0.55$. The rationale behind this is that certain application classes, such as VR and video, demonstrate very similar behavior, making it challenging to distinguish between them. A different results are observed when classifying application classes, see Fig.~\ref{fig:class}. Most algorithms find it easier to differentiate and correctly classify samples in this context. Notably, random forest performs best, closely followed by XGBoost. This outcome supports our hypothesis that there is no need to resort to computationally expensive algorithms like XGBoost for classifying application categories. Simpler algorithms, such as random forest, can yield even better results. However, we note that as compared to Mann-Whitney test, 80\% accuracy is achieved by utilizing 50\% out of 120 beam tracking intervals implying that the detection delay after application initialization is around 5 s.

\begin{figure}
\centering
\includegraphics[width=0.65\columnwidth]{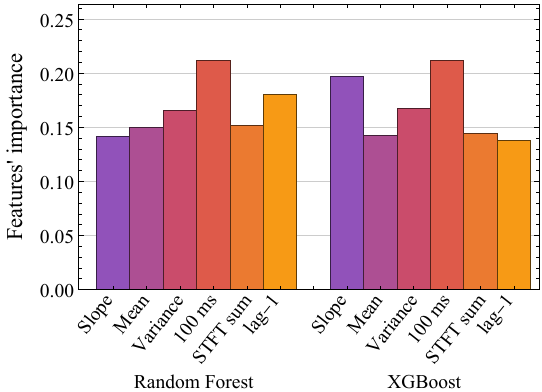}
\caption{Features' importance for Random Forest and XGBoost classifiers.}
\label{fig:featureimportance}
\end{figure}

Finally, Fig.~\ref{fig:featureimportance} compares the feature importance for the two leading algorithms. For both algorithms, the 100 ms value is highly significant. This implicitly highlights the efficiency of the conventional statistical approach based on Mann-Whitney test. However, random forest places a strong emphasis on autocorrelation, whereas for XGBoost, autocorrelation is the least important feature.

%\textcolor{red}{Anna, can you please, add here: (i) utilized specific ML classifiers, (ii) short description of tables with values of features as well as data representation at the input of the ML algorithms, (iii) short description of their parameters and their choice, (vi) add figures: three in a row for applications level classification (accuracy, recall, and F1) and three for binary classification between classes (also accuracy, recall, and F1), and (v) their description.}

\section{Conclusions}\label{sect:conc}

Similarly to dynamic human body blockage, micromobility causes large and quick deviations in the received signal power that may eventually lead to outage even when regular beam tracking is utilized. One the other hand, for some applications when the user is stationary micromobility impairments are non-noticeable barely affecting the signal received power allowing to increase the time interval between beam tracking time instants saving power of the handheld devices.

In this paper, for the first time, we carried our an extensive measurements campaign of micromobility process in the sub-THz frequency band at 156 GHz. We reported measurements and revealed their statistical characteristics showing that conventional applications are characterized by drastically different behavior of the received signal strength after the beam tracking time instant. We then proceeded proposing two algorithms for on-line detection of the application and their classes. For conventional statistical based tests we demonstrated that one may infer application class not the specific application within just 1 s after application initialization. By utilizing the ML techniques we demonstrated that the highest achievable accuracy is just 80\% for random forest and XGBoost algorithms for longer time interval after application initialization (5 s of more). Nevertheless, for accurate detection of the application classes based on the received signal strength we recommend utilizing both approaches simultaneously as they are not computationally intensive and operate using the same set of statistical data.

\section*{Acknowledgements}
The publication was supported by the grant for research centers in the field of AI provided by the Analytical Center for the Government of the Russian Federation (ACRF) in accordance with the agreement on the provision of subsidies (identifier of the agreement 000000D730321P5Q0002) and the agreement with HSE University No70-2021-00139.

\bibliographystyle{ieeetr}
\bibliography{main}  %%% Uncomment this line and comment out the ``thebibliography'' section below to use the external .bib file (using bibtex) .

\end{document}